# In-Situ Monitoring of the Charge Carrier Dynamics of $CH_3NH_3PbI_3$ Perovskite Crystallization Process

*Efthymis Serpetzoglou[1,3], Ioannis Konidakis[1], Temur Maksudov[2,4], Apostolos Panagiotopoulos[2,4], Emmanuel Kymakis[2], Emmanuel Stratakis[1,4,*]*

[1]*Institute of Electronic Structure and Laser (IESL), Foundation for Research and Technology-Hellas (FORTH), 71110, Heraklion, Crete, Greece*

[2]*Center of Materials Technology and Photonics, Electrical Engineering Department, Technological Educational Institute (TEI) of Crete, 71004, Heraklion, Crete, Greece*

[3]*Department of Physics, University of Crete, Herakleio, Greece*

[4]*Department of Materials Science and Technology, University of Crete, Greece, 71003, Heraklion, Crete, Greece*

**Corresponding Author**

*+30 2810-391274. E-mail: stratak@iesl.forth.gr

**Abstract**

Although methylammonium lead iodide ($CH_3NH_3PbI_3$) perovskite has attracted enormous scientific attention over the last decade or so, important information on the charge extraction dynamics and recombination processes in perovskite devices is still missing. Herein we present a novel approach to evaluate the quality of $CH_3NH_3PbI_3$ layers, via in-situ monitoring of the perovskite layer charge carrier dynamics during the thermal annealing crystallization process, by means of time-resolved femtosecond transient absorption spectroscopy (TAS). In particular, $CH_3NH_3PbI_3$ films were deposited on two types of polymeric hole transport layers (HTL), poly(3,4-ethylenedioxythiophene)-poly-(styrenesulfonate) (PEDOT:PSS) and poly-(triarylamine) (PTAA), that are known to provide different carrier transport characteristics in perovskite solar cells. In order to monitor the evolution of the perovskite charge carrier dynamics during the crystallization process, the so-formed $CH_3NH_3PbI_3$/HTL architectures were studied in-situ by TAS at three different annealing temperatures, i.e. 90, 100 and 110 ºC. It is revealed that the annealing time period required in order to achieve the optimum perovskite film quality in terms of the decay dynamics strongly depends on the annealing temperature, as well as, on the employed HTL. For both HTLs, the required period decreases as higher annealing temperature is used, while, for the more hydrophobic PTAA polymer, longer annealing periods were required in order to obtain the optimum charge carrier dynamics. The correlation of the TAS finding with the structural and morphological features of the perovskite films is analysed and provides useful insights on the charge extraction dynamics and recombination processes in perovskite optoelectronic devices.

**Introduction**

During the last decades organic-inorganic lead halide perovskites have attracted tremendous scientific attention due to their unique optical properties and their potential employment in various types of state-of-the-art optoelectronic applications such as next-generation photovoltaics, light-emitting devices, photodetectors, random lasers, and light-emitting diodes.[1-5] Among others, the important properties that render perovskite films a suitable light absorber component of solar cell devices, are the medium optical band gap and strong absorption coefficients, the long carrier diffusion lengths as well as, the low recombination losses and band gap tunability.[6-9] Remarkably, over the last decade, the power conversion efficiency (PCE) of the lead halide perovskite solar cells (PSCs) has been boosted from ~4% to over ~22%.[10-13] Despite this impressive increase in the PCE, there is still significant lack of understanding of the charge extraction processes as well as the recombination dynamics which significantly limit the efficiency attained, compared to the theoretical predictions.[14-17]

Several approaches have been adopted in order to tackle the aforementioned scientific challenges. Many studies focus on the perovskite synthesis procedure itself, aiming to optimize its crystalline quality, as it is generally proven that the quality of the absorber plays a crucial role on the device performance.[11,12,18-20] Along these lines, the importance of various types of one- and two-step deposition procedure, single-source vapour and fully-vacuum deposition, inject-printing and resonant infrared matrix-assisted pulsed laser evaporation methods has been explored,[19,21-25] while the advantages of employing rapid laser-assisted annealing crystallization techniques have also been demostrated.[26-28]

Another important aspect towards improving the device efficiency, is the selection of suitable hole transport (HTL) and electron transport (ETL) polymer layers. In particular, the interactions between HTL/perovskite and perovskite/ETL interfaces are known to play a crucial role on the device PCE. Based on this finding, several efforts have been made in order to mitigate the factors that reduce the PCE and cause anomalous hysteresis response, such as ion migration in the crystalline lattice of the perovskite and the formation of light activated meta-stable charge states.[29-31] Aiming to resolve these issues, recent studies report on the doping of the hole and electron transfer layers.[32-35] Furthermore, many groups focus on the effect of transfer materials (HTL and ETL) on the electrical characteristics and stability of the devices.[12,13,36-39]

Herein, the perovskite charge carrier dynamics are monitored in-situ, during the crystallization process, by means of transient absorption spectroscopy (TAS). This approach enabled us identify the optimum annealing time period for different HTL/perovskite configurations (i.e. PEDOT:PSS/$CH_3NH_3PbI_3$ and PTAA/$CH_3NH_3PbI_3$). The in-situ monitoring of the TAS components revealed crucial information for the carrier transport dynamics, which critically affect the overall performance of the respective electronic devices. To date, the optimum annealing period for the crystallization process was determined in an empirical manner, via the evolution of the perovskite XRD patterns or only after completing the solar cell fabrication, via measuring the photovoltaic efficiency. This work first shows an alternate approach to optimize the perovskite layer crystallization process, based on the in-situ measurement of charge transport properties of HTL/perovskite configurations. In particular, the charge extraction properties were investigated at three different thermal annealing temperatures applied during the TAS measurement, i.e. 90 ºC, 100 ºC and 110 ºC, while employing two different HTL polymers, i.e. the more hydrophilic, PEDOT:PSS, and the more hydrophobic, PTAA. All TAS measurements were performed entirely under inert conditions within a specially designed homemade cell to exclude any ambient-induced degradation effects. It is well known that the charge carrier dynamics and recombination rates, probed by TAS, are strongly correlated with the electrical performance of PSC devices[14,21,35,41,42,48] and also, the nature of the employed HTL polymer affects the charge extraction processes and thus the photovoltaic efficiency.[7,14,21,28,36,40-51]

Particularly, we observe that in case of PTAA/$CH_3NH_3PbI_3$ architectures, the charge carrier extraction and injection processes are significantly faster compared to the corresponding time components of the PEDOT:PSS/$CH_3NH_3PbI_3$ system, i.e. regardless the selected crystallization temperatures. Moreover, we found that the annealing time period required to achieve optimum perovskite film quality, in terms of the decay dynamics, strongly depends on the crystallization temperature, as well as, on the HTL. For both HTLs, the required annealing time period decreases as a higher annealing temperature is used, whereas strikingly, for the more hydrophobic PTAA polymer, longer annealing periods are required in order to reach the optimum charge carrier dynamics. The TAS findings are correlated with the structural and morphological characteristics of the so-formed perovskite films, investigated by x-ray diffraction (XRD), scanning electron microscopy (SEM) and absorption spectroscopy. As a consequence, insights on carrier transport properties and extraction dynamics are acquired, which are useful towards the fabrication of high-quality perovskite layers for optoelectronic devices.

**Experimental Section**

*Perovskite Film Preparation*

Perovskite films were prepared on 20 x 15 mm pre-patterned indium tin oxide (ITO) coated glass substrates (Naranjo Substrates) and a sheet resistance of ~20 Ω sq$^{-1}$. A three-step cleaning procedure with deionized water, acetone, isopropanol was followed to clean the substrates from any impurities. Following this, the substrates were transferred inside an ultraviolet ozone cleaner in order to increase the hydrophilicity on the surface of ITO/glass substrate and to remove organic residues. Subsequently, the ITO/glasses were ready for the deposition of the PEDOT:PSS or PTAA HTL polymers.

For the samples where PEDOT: PSS was used as HTL, the polymer was spin coated on the ITO surface from aqueous solution (4000 rpm for 60 s), followed by a thermal annealing at 120 °C for 15 minutes in ambient conditions and 30 more minutes inside a nitrogen-filled glovebox at 120 °C. A result of this spin-coated process was a ~30 nm layer thickness.[21,28,35] For samples with the other employed HTL polymer, a solution of PTAA powder in toluene (7mg ml$^{-1}$) doped with 1.5 % wt tetrafluoro-tetracyanoquinodimethane was prepared and spin coated (4000 rpm for 35 s) on the ITO surface and annealed at 110 °C for 10 minutes. After the HTL deposition in both cases, the substrates were left to cool down to room temperature (RT).

Perovskite film fabrication took place inside the nitrogen-filled glove box. The methylammonium lead iodide (CH$_3$NH$_3$PbI$_3$) films were prepared by a typical solution-based two-step procedure. For this purpose, two anhydrous solution were prepared separately. The first solution was prepared by dissolving lead iodide (PbI$_2$) in dimethylformamide (DMF) and stirred overnight at 75 °C, while the second one, by dissolving methylammonium iodide (CH$_3$NH$_3$I) in isopropanol and stirring at room temperature. In step one, the solution (PbI$_2$) (450mg ml$^{-1}$) was spin coated on the HTL and heated at 100 °C for 10 minutes. For the thermal annealing configurations, in step two the CH$_3$NH$_3$I solution (45mg ml$^{-1}$) was spin coated on top of the surface of PbI$_2$ and heated immediately at 100 ºC for 30 minutes. This was the final step for the preparation of HTL/CH$_3$NH$_3$PbI$_3$ architectures synthesized upon following the typical thermal annealing (TA) crystallization procedure (TA samples). Rather differently, for the corresponding samples prepared for the in-situ TAS (IS samples), after the CH$_3$NH$_3$I solution was casted (step 2), the samples remained at RT for 30 minutes in inert conditions, before heated at the desired temperatures of 90, 100 and 110 ºC, with a heating rate of 20 ºC/minute, while TAS measurements were performed every 10 minutes after the selected annealing temperature was reached. For the sake of convenience, all crystallization parameters for the fabricated HTL/CH$_3$NH$_3$PbI$_3$ samples are summarized in Table 1. In one of our previous studies[21] the thickness of the perovskite layer for typical TA samples was found 350 ± 50 nm when PEDOT:PSS is used and 450 ± 50 nm when PTAA is employed. For this work, we have now measured and show in the revised supporting information (Figure S1) indicative cross-section SEM images for IS samples at 100 ºC. Notably, the perovskite films thickness of these IS samples, are found to be similar for both PEDOT:PSS and PTAA configurations with the corresponding TA samples, i.e. 354 nm and 452 nm. Thus, it appears that the employed crystallization procedure, i.e. TA or IS, does not affect significantly the perovskite film thickness.

*Characterization techniques and measurements*

The crystal structure of the CH$_3$NH$_3$PbI$_3$ perovskite was studied by an X-Ray Rigaku (RINT-2000) Diffractometer operating with a continuous scan of Cu Ka1 radiation with λ = 1.54056 Å. The scan rate for all XRD measurements was 0.1º/s in the range of 2θ = 5º - 60º. The absorption spectra on the wavelength range of 470-900 nm, were recorded by a PerkinElmer (Lamda950)

UV/VIS Spectrometer. The morphology of the perovskite was examined by JEOL, JSM-7000F field emission scanning electron microscope. The roughness of the two studied HTLs, i.e. PEDOT:PSS and PTAA polymers, were measured by a Park XE-7 Atomic Force Microscopy (AFM) instrument in tapping mode. The scan rate for the AFM measurements was 0.3 Hz, while the total area was 1x1 μm. Finally, a PerkinElmer UV/vis (Lambda 950) spectrometer was used to measure the absorption of the fabricated HTL/$CH_3NH_3PbI_3$ films over the wavelength range of 430−900 nm.

Transient Absorption Spectroscopy (TAS) measurements were performed on a Newport (TAS-1) transient absorption spectrometer. The light source was a pulsed laser from an Yb:KGW laser system (PHAROS, Light Conversion) with central wavelength at 1026 nm, pulse duration 170 fs, and repetition rate of 1 KHz. As depicted in Figure S2, the fundamental beam (1026 nm) was split at the pump beam (90% of the initial beam) and the probe beam (10% of the initial beam). It should be mentioned that the excitation of the active layer (perovskite layer) occurred via a two-photon absorption process. The sum of the photon energies corresponding to two, 1026 nm, photons (i.e. ~2.42 eV), exceeds the band gap of $MAPbI_3$ (~1.6 eV). The pump beam was used for the sample excitation. The other part of the fundamental beam, the probe beam, was passed through the delay line (up to 8 ns) and then was focused on a non-linear crystal (YAG crystal), generating a supercontinuum white light of 530 – 970 nm. The energy of both beams can be controlled by a reflective neutral density filter. The probe beam, after the sample, was coupled through an optical fiber and monitored as a function of wavelength at various delay time after photoexcitation.

**Results and Discussion**

*Structural and Morphological characterization*

It is well known that the PCE of the PSCs is strongly correlated with the crystalline properties of the perovskite absorber and is notably affected by the grain boundaries, grain size and the perovskite phase.[9] Figure 1 presents the XRD patterns for the different IS perovskite layers deposited onto the two HTLs used here. All patterns are dominated by two main peaks at 14° and 28.4°, corresponding to the (110) and (220) lattice planes of $CH_3NH_3PbI_3$, respectively.[36,52-55] The much weaker peak at 31.8° is attributed to the (310) lattice plane, while the (220) crystal reflection that is expected at 24.4° is not observed. Also, the peak at 12.5°, corresponding to an excess $PbI_2$, is present in all XRD graphs. Some studies have already shown that the residual or unreacted $PbI_2$ can decrease the surface states and defects in the perovskite layer.[20,56,57] Furthermore, the slightly over-stoichiometric $PbI_2$ can enhance the electrical characteristics and the electric quality of the PSCs through the passivation effect.[40,57,58] It can be concluded that all XRD profiles of the perovskite layers formed on the two HTLs used, exhibit the same peaks, however as the thermal annealing temperature rises the intensity of $PbI_2$ peak marginally increases, while the intensity of the peak corresponding to (110) lattice plane at 14° slightly decreases for PEDOT:PSS/$CH_3NH_3PbI_3$ configuration, while increases for PTAA/$CH_3NH_3PbI_3$ architecture. It is also noted that, regardless the temperature, the IS samples of the PEDOT:PSS/$CH_3NH_3PbI_3$ system present weaker $PbI_2$ peaks compared to the corresponding TA ones (Figure S3),[21] while the PTAA/$CH_3NH_3PbI_3$ IS samples always show a stronger $PbI_2$ peak (Figure S3).[21] The $CH_3NH_3PbI_3$/$PbI_2$ peak ratios for all IS samples are summarized in Table S1 and it is clear that the highest ratio for PEDOT:PSS/$CH_3NH_3PbI_3$ configuration is 2.83 and is observed at 90 ºC, while for PTAA/ $CH_3NH_3PbI_3$ architecture is 1.63 and is observed at 110 ºC.

In order to explore the morphological differences of the various perovskite layers prepared, SEM imaging presented in Figure 2, was employed. In particular, Figures 2a,e show the SEM images of the TA perovskite layers grown onto the PEDOT:PSS and PTAA, respectively, whereas Figures 2b-d and Figures 2f-h illustrate the SEM images for the IS samples at all studied temperatures. It should be mentioned that the XRD patterns as well as the SEM images for IS samples, were captured following the annealing time period that the optimum decay time components were observed, as it will be described below.

In agreement to previous findings,[21,28] inspection of Figures 2a and 2e shows that, in the case of TA samples, a larger perovskite grain size is obtained for the PTAA architectures. However, when one considers the IS samples, the perovskite grain size in case of PEDOT:PSS polymer is larger than the one grown on PTAA. Moreover, in the case of PEDOT:PSS/$CH_3NH_3PbI_3$ system, the perovskite grain size for the TA sample (Figure 2a) is always smaller when compared to that of IS samples, regardless the annealing temperature; in particular, larger grains are obtained at 90 ºC. Rather differently, in the case of PTAA/$CH_3NH_3PbI_3$ IS samples the perovskite grains are larger for the two higher in-situ annealing temperatures i.e. 100 ºC and 110 ºC, compared to the grain size of the TA sample. This can be attributed to the slightly different thermal crystallization history among the TA and the IS samples employed for the in-situ TAS studies.

The corresponding AFM images of the PEDOT:PSS and PTAA polymer surfaces are shown in Figures S4a,b. The average roughness values estimated from these images are equal to 0.942 and 0.451 nm for the PEDOT:PSS and PTAA, respectively. Furthermore, in Figures S4c,d one can see that the absorbance spectra exhibit the two distinct features at 490 nm and 760 nm attributed to optical transitions within the dual valence band structure of $CH_3NH_3PbI_3$.[7,36,41-43] For all studied IS samples as well as for the TA samples for each employed HTL polymer, there is no noticeable shift for the two characteristics absorption features (Figures S4c,d). Notably, in the case of more hydrophilic PEDOT:PSS the absorption intensity of both TA and IS sample is approximately equal, while when the more hydrophobic PTAA polymer is employed, the intensity for TA sample is significantly higher compared with the corresponding IS samples. This is additional evidence, that the in-situ crystallization is not favored upon using the more hydrophobic PTAA polymer.

*TAS Studies*

Understanding the carrier transport properties at the HTL/perovskite interfaces is considerably important towards the enhancement of the performance of PSCs. It is well known that TAS is a powerful tool in order to extract information regarding charge carrier dynamics and recombination processes.[7,14,21,28,36,40-51] In order to shed light on the crystallization process of both PEDOT:PSS/CH$_3$NH$_3$PbI$_3$ and PTAA/CH$_3$NH$_3$PbI$_3$ configurations, we performed in-situ TAS measurement at three different annealing temperatures, i.e. 90, 100 and 110 °C. It is necessary to mention that the TAS measurements for the reference TA samples were performed at room temperature, while in-situ measurements in the IS samples were obtained at the aforementioned three temperatures following crystallization.

Figure 3 displays typical TAS spectra of delta optical density (ΔOD) as a function of wavelength at t=0 ps for the IS samples formed at the three studied temperatures, as well as for the TA samples, following photoexcitation at 1026 nm with a pump fluence of 1.5 mJcm$^{-2}$. ΔOD is the change in absorption of the perovskite layer after photoexcitation and is calculated as ΔOD = log(blocked/unblocked) = OD$_{probe}$ − OD$_{pump+probe}$. The main ΔOD peak at ~740 nm is attributed to the transient photo-induced bleaching of the band edge transition, while a photo-induced transient absorption (PIA) in the range of 530-700 nm is additionally observed.[7,21,28,36] As we can see in Figure 3, there is a blue shift of the ΔOD peak for all IS samples compared to the reference TA ones, attributed to the different thermal annealing procedures used for the IS and TA sample formation. Frost et al. have already shown the dependence of the $CH_3NH_3PbI_3$ bandgap on temperature changes. In particular, the bandgap of the perovskite increases as the lattice enlarges due to the increase of temperature.[59] This is the result of the splitting of the bonding and antibonding orbitals of $CH_3NH_3PbI_3$. Thus, the thermal extension of the lattice, due to temperature rise, affects the conduction band minimum as well as the valence band maximum leading to an overall bandgap increase.

The time-resolved relaxation dynamics were thoroughly investigated following both an exponential[21,28,35,42,44,48-50] and polynomial fitting analysis.[14,21,28,40,41,43,45-47] From the exponential fitting analysis we have the ability to determine the critical time components of the charge carrier transport processes between the perovskite film and the employed HTL polymers, while the polynomial one provides information on the kinetic rates for the charge carrier dynamics processes that occur in the perovskite film. Figure 4 displays the typical ΔOD versus wavelength plots at various time delays for the two studied configurations at 100 °C, and the corresponding three-exponential fittings based on the equation $y = y_0 + A_1 \exp(-x/\tau_1) + A_2 \exp(-x/\tau_2) + A_3 \exp(-x/\tau_3)$. Furthermore, Figure 5a and Figure 5b depict the transient band edge bleach kinetics and the corresponding decay three-exponential fittings at 90 °C (left) and 110 °C (right) for PEDOT:PSS/$CH_3NH_3PbI_3$ and PTAA/$CH_3NH_3PbI_3$ configurations, respectively. The resolution of each measurement is 1 ps, but for the sake of clarity for the presentation, we used skip point 12 for all relaxation plots. All kinetic fit parameters for the two studied systems, as obtained at each temperature, are summarized in Tables 2 and 3.

In particular, the fast time component ($\tau_1$) is attributed to the charge carrier trapping at the perovskite grain boundaries and perovskite/HTL interfaces[21,28,36] and not generally to the perovskite defects.[60,61] It represents the time required for the transition of an excited electron from the conduction band to the trap states. Smaller $\tau_1$ times indicate quicker traps filling, leading to larger splitting in the quasi-Fermi energy levels resulting in the enhanced open-circuit Voltage ($V_{OC}$) of the device. The $V_{OC}$ is defined as the difference between the quasi-Fermi energies of electrons and holes in the solar cell under illumination. The quicker traps filling and the more efficient free charge carrier injection lead to considerably enhanced electrical characteristics of the perovskite films.[62,63] It's worth mentioning that $\tau_1$ component is not only affected by the density of trap states, but also from their depth as well.[64] Notably, for the PEDOT:PSS/CH$_3$NH$_3$PbI$_3$ system at 90 °C, $\tau_1$ decreases from 45 ps after an annealing time of 25 min of to 22 ps after 85 min and then increases again to 25 ps after 95 min where it stabilizes (Table 2). Similarly, for the sample annealed at 100 °C $\tau_1$ drops from 29 ps after an annealing time of 15 min to 22 ps after 55 min and then rises up to 35 ps after 85 min (Table 2). Accordingly, $\tau_1$ for sample annealed at 110 °C drops from 29 ps after an annealing time of 15 min to 22 ps after 25 min, and finally, stabilizes at 24 ps after 35 min (Table 2). These results reveal that, while the annealing period progresses, $\tau_1$ decreases until the optimum (faster) value of 22 ps, which is indicative of the best quality perovskite film, and then increases again suggesting that further annealing is not favorable for the crystalline quality of the CH$_3$NH$_3$PbI$_3$ films.

The second-time component ($\tau_2$) represents the time required for a hole injection from the perovskite layer to the HTL polymer including the corresponding time for the hole diffusion to the perovskite/HTL interface.[21,28,36] It is well-known that the faster the hole injection the better the electrical characteristics of the respective devices.[48,65] According to our results, at 90 °C, $\tau_2$ drops from 185 ps after annealing time of 25 min to 105 ps after 85 min and then increases to 276 ps after 145 min of total annealing time. At 100 °C $\tau_2$ decreases from 321 ps after annealing time of 15 min to 190 ps after 55 min, respectively, and then rises to 303 ps after 85 min of annealing. The same trend is observed at 110 °C, as $\tau_2$ decreases from 257 ps to 210 ps after annealing time of 25 min and then increases to 257 ps after 55 min. Similar to $\tau_1$, $\tau_2$ is following the same trend and reaches its optimum (lowest) value at the same annealing time period for the three studied temperatures. These annealing time periods are identified equal to 85, 55, and 25 min for the annealing temperatures of 90 °C, 100 °C, and 110 °C, respectively.

It should be noted that the IS PEDOT:PSS/CH$_3$NH$_3$PbI$_3$ samples exhibit significant faster $\tau_2$ time components compared to the $\tau_2$ of the TA samples (Table 2).[21] Thus, the IS samples are expected to exhibit potentially better electrical characteristics that the TA ones, which complies with the observed higher relative peak intensity ratio CH$_3$NH$_3$PbI$_3$/PbI$_2$ in the XRD patterns of Figure 1 (Figure S3). It is well known that the excess of PbI$_2$ is harmful for the composition and optical properties of the perovskite film, while it can be beneficial for the electrical performance.[20,54,56-58] Table 2 also lists the third long-life time component ($\tau_3$), which is representative of the exciton recombination time.[21,28,36] For all studied temperatures and annealing periods, the $\tau_3$ time component was found to be of the same order of magnitude, i.e. decades of μs.

In the case of PTAA/CH$_3$NH$_3$PbI$_3$ system at 90 °C $\tau_1$ decreases from 21 ps after annealing time of 25 min to 17 ps after 85 min and then increases to 29 ps after 145 min, while at 100 °C $\tau_1$ remains constant at around 21 ps (Table 3). Moreover, $\tau_1$ at 110 °C drops from 50 ps to 19 ps after annealing time of 25 min and 35 min, respectively, and then increases to 51 ps after 115 min (Table 3). While, $\tau_2$ at 90 °C, drops from 69 ps after annealing time of 25 min to 48 ps after 85 min and then increases to 110 ps after 145 min. At 100 °C $\tau_2$ decreases from 85 ps after annealing time of 35 min to 60 ps after 75 min, and then rises to 102 ps after 105 min. A similar trend is observed at 110 °C, where $\tau_2$ decreases from 522 ps to 63 ps after annealing time of 35 min and then increases to 510 ps after 115 min. Therefore, for the PTAA/CH$_3$NH$_3$PbI$_3$ system, the optimum annealing time periods for both $\tau_1$ and $\tau_2$ time components are found to be the same regardless the annealing temperature. Besides this, the optimum annealing time periods are revealed to be longer compared to the corresponding periods for the PEDOT:PSS/CH$_3$NH$_3$PbI$_3$ system. Thus, for the case of PTAA polymer longer annealing periods are required in order to achieve optimum charge transfer dynamics, at least for the two higher crystallization temperatures of 100 °C and 110 °C. Namely, the optimum annealing time periods are identified equal to 85, 75, and 35 min for the annealing temperatures of 90 °C, 100 °C, and 110 °C, respectively. Finally, Table 3 includes the $\tau_3$ time component (recombination time) for PTAA/perovskite configuration. For all studied temperatures and annealing time periods, this time component was found to be of the same order of magnitude (hundreds of picoseconds).

It should be emphasized that at the optimum annealing time periods, the PTAA/CH$_3$NH$_3$PbI$_3$ IS samples exhibit faster $\tau_2$ time component compared to that of the TA samples (Table 3). Such faster charge carrier dynamics are accompanied by a significantly lower CH$_3$NH$_3$PbI$_3$/PbI$_2$ XRD peak ratio (Figures 1 and S3 and Table S1). Therefore, in the case of the more hydrophobic PTAA polymer, the in-situ TAS annealing process results to excessive presence of residual PbI$_2$.

We perform also the well-established polynomial fitting model based on the rate equation $dn(t)/dt = - k_3n^3 - k_2n^2 - k_1n$, where n is the charge carrier density and $k_1$, $k_2$ and $k_3$ are the trap-assisted recombination rate, bimolecular recombination rate and Auger trimolecular recombination rate, respectively (Figure S5).[14,40,41,43,45-47,51] In a recent paper, Wehrenfennig et al.[41] have reported on the correlation between the $k_2$ bimolecular recombination rate and the efficiency of the PSCs. It is particularly proposed that slower $k_2$ rates are indicative of longer free-charge carrier diffusion lengths, that consequently result in better performance of the planar heterojunction devices. Table 4 lists the kinetic rates for the PEDOT:PSS/CH$_3$NH$_3$PbI$_3$ system for the TA and IS samples at the optimum crystallization annealing time period, while Table 5 includes the corresponding rates for the PTAA/CH$_3$NH$_3$PbI$_3$ system. It is clear that all the IS samples exhibit smaller $k_2$ rates compared to those of TA ones. Moreover, the slower bimolecular recombination rates for both configurations occurs at 90 ºC, which complies with the faster $\tau_2$ time component of the PEDOT:PSS/CH$_3$NH$_3$PbI$_3$ and PTAA/CH$_3$NH$_3$PbI$_3$ systems (Tables 2 and 3). The above findings are also in agreement with the larger grains size and the higher CH$_3$NH$_3$PbI$_3$/PbI$_2$ peak ratio obtained from the XRD patterns in case of PEDOT:PSS/CH$_3$NH$_3$PbI$_3$ at 90 ºC. On the contrary, in the case of PTAA/CH$_3$NH$_3$PbI$_3$, the larger grains size and the higher CH$_3$NH$_3$PbI$_3$/PbI$_2$ peak ratio are observed at 110 ºC, while the faster carrier injection and slower bimolecular recombination rate are observed at 90 ºC. These results indicate that for the PTAA/CH$_3$NH$_3$PbI$_3$ configuration, the perovskite layer crystallinity and the corresponding charge carrier dynamics are more affected by the thermal crystallization history, compared to the PTAA/CH$_3$NH$_3$PbI$_3$ configuration.

In summary, it was found that, all IS samples for both configurations tested exhibited faster charge carrier dynamics and slower recombination rates compared with the corresponding TA samples. Namely, in the case of PEDOT:PSS/CH$_3$NH$_3$PbI$_3$ the faster charge carrier dynamics and slower recombination rates are obtained at 90 ºC which is in agreement with the best obtained crystalline quality of the perovskite film as revealed from the XRD pattern (highest $CH_3NH_3PbI_3/PbI_2$ peak ratio), and larger perovskite grain sizes as emerged from SEM studies. On the contrary, in the case of PTAA/$CH_3NH_3PbI_3$ architecture, the highest $CH_3NH_3PbI_3/PbI_2$ peak ratio and larger perovskite grain sizes are observed are 110 $^oC$, whereas the faster carrier dynamics and slower recombination rates are exhibited at 90 $^oC$. Based on the above it is revealed that the different thermal history of the in-situ (IS) TAS crystallization process does not affect the formation of the perovskite film upon employing the more hydrophilic PEDOT:PSS polymer as HTL. Rather differently, a discrepancy is noticed in the case of PTAA/perovskite configuration as the optimum charge extraction processes and crystalline morphology are obtained at different temperatures. It is worth to stress out, that this effect is not observed when the typical thermal annealing route is followed for the same architectures. Furthermore, when the more hydrophilic PEDOT:PSS is employing as HTL, faster annealing periods are required in order to achieve the most efficient charge extraction properties at each temperature, compared with the corresponding period for PTAA/$CH_3NH_3PbI_3$ architecture.

**Conclusions**

By means of TAS we monitored in-situ the evolution of charge extraction dynamics of $CH_3NH_3PbI_3$ films throughout the crystallization process that took place at three different thermal annealing temperatures. The perovskite films were crystallized on the surface of two well-studied HTL polymers, the hydrophilic PEDOT:PSS and the more hydrophobic PTAA, in order to explore also the HTL surface wettability effect on the perovskite crystallization and the corresponding charge extraction properties. It was found that the different crystallization procedure and different thermal history for the in-situ (IS) samples seem to favor the formation of the perovskite layer deposited onto the more hydrophilic PEDOT:PSS polymer. On the contrary, when the more hydrophobic PTAA polymer is used, the in-situ crystallization routes disrupt the crystallization of the perovskite film. As a consequence, longer annealing periods are required in order to achieve the most efficient charge extraction properties. Therefore, the duration of thermal annealing for developing perovskite films with optimum charge extraction dynamics strongly depends on the surface nature of the employed HTL polymer. Our results reveal that in-situ TAS is a powerful technique and constitutes a figure-of-merit diagnostic tool towards the identification of the best quality perovskite film (optimum crystallization period at specific temperature).

ASSOSIATED CONTENT

Supporting Information

Experimental details, TAS setup, characterization of materials (X-ray Diffraction, Atomic Force Microscopy – AFM for HTL polymers, Absorption Spectroscopy).

AUTHOR INFORMATION

**Corresponding Authors**

*E-mail: stratak@iesl.forth.gr. Phone: +30 2810-391274 (E.S.).

ORCID

Efthymis Serpetzoglou: 0000-0002-5218-023X
Ioannis Konidakis: 0000-0002-2600-2245
Emmanuel Kymakis: 0000-0003-0257-1192
Emmanuel Stratakis: 0000-0002-1908-8618

**Notes**

The authors declare no competing financial interest.

ACKNOWLEDGMENT

The authors are grateful to A. Manousaki and K. Savva (all in IESL, FORTH), and M. Krassas (TEI) for their technical assistance with SEM, XRD and AFM studies respectively. This work is supported by the European Research Infrastructure NFFA-Europe, funded by EU's H2020 framework program for research and innovation under grant agreement number 654360.

**Figures and Tables**

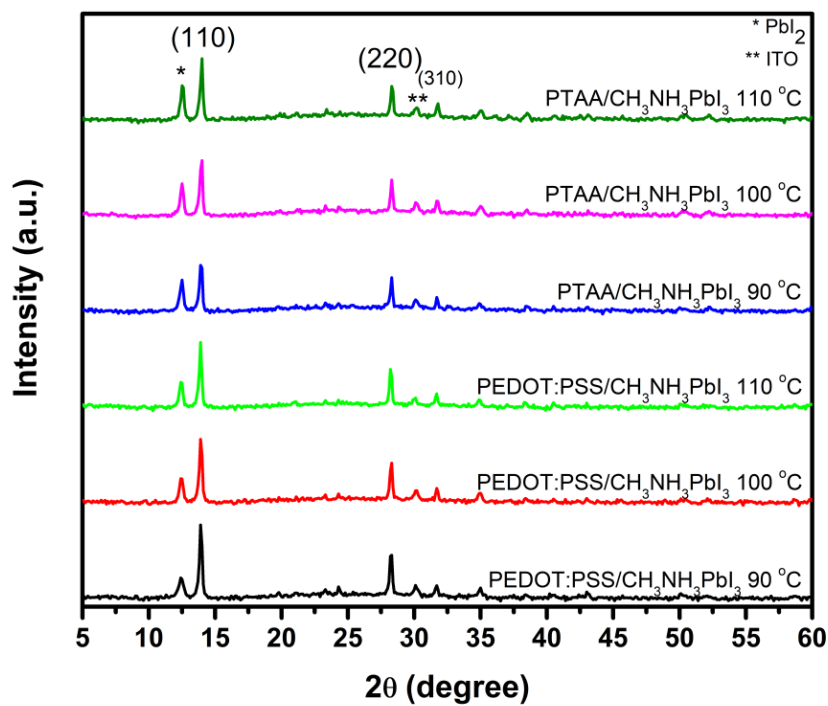

**Figure 1:** XRD patterns of the PEDOT:PSS/CH$_3$NH$_3$PbI$_3$ and PTAA/CH$_3$NH$_3$PbI$_3$ architectures at all studied temperatures.

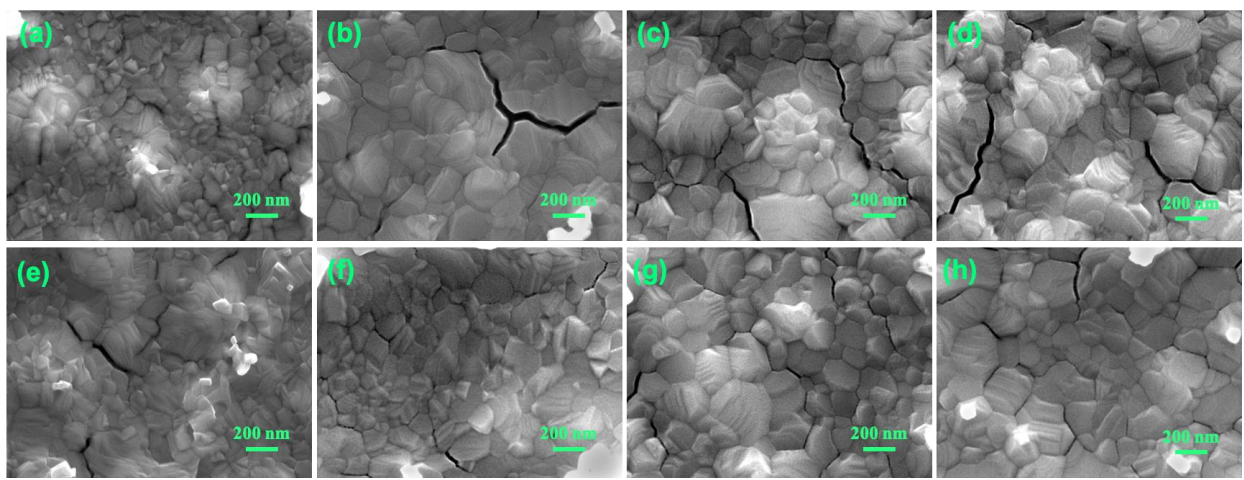

**Figure 2:** SEM images for the PEDOT:PSS/CH$_3$NH$_3$PbI$_3$ configuration thermal annealing (a) at 90 ºC (b), 100 ºC (c), 110 ºC (d) and for PTAA/CH$_3$NH$_3$PbI$_3$ architecture thermal annealing (e) at 90 ºC (f), 100 ºC (g) and 110 ºC (h), respectively.

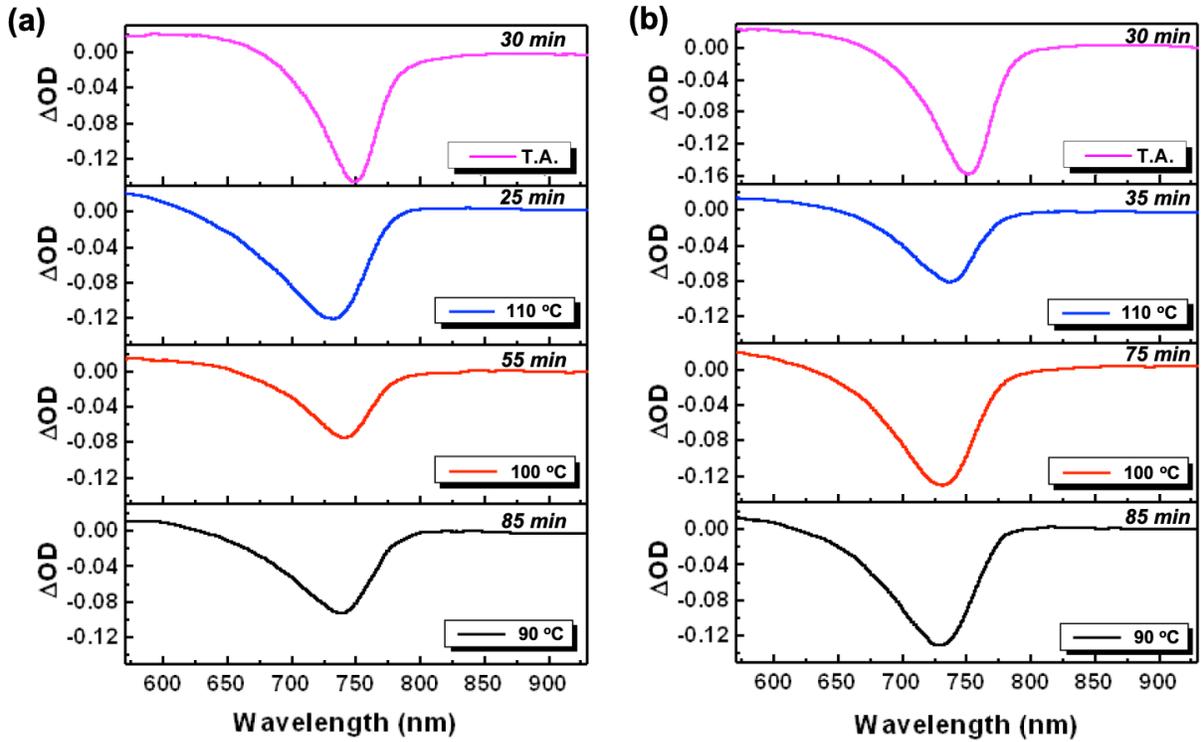

**Figure 3:** Typical TAS spectra of ΔOD as a function of wavelength at t=0 ps, for 90, 100 and 110 ºC and Reference (Thermal Annealing at 100 ºC on hot plate − TA) for Glass/ITO/PEDOT:PSS/CH$_3$NH$_3$PbI$_3$ (a) and for Glass/ITO/PTAA/CH$_3$NH$_3$PbI$_3$ (b) configurations following photoexcitation at 1026 nm with a pump fluence of 1.5 mJ cm$^{-2}$.

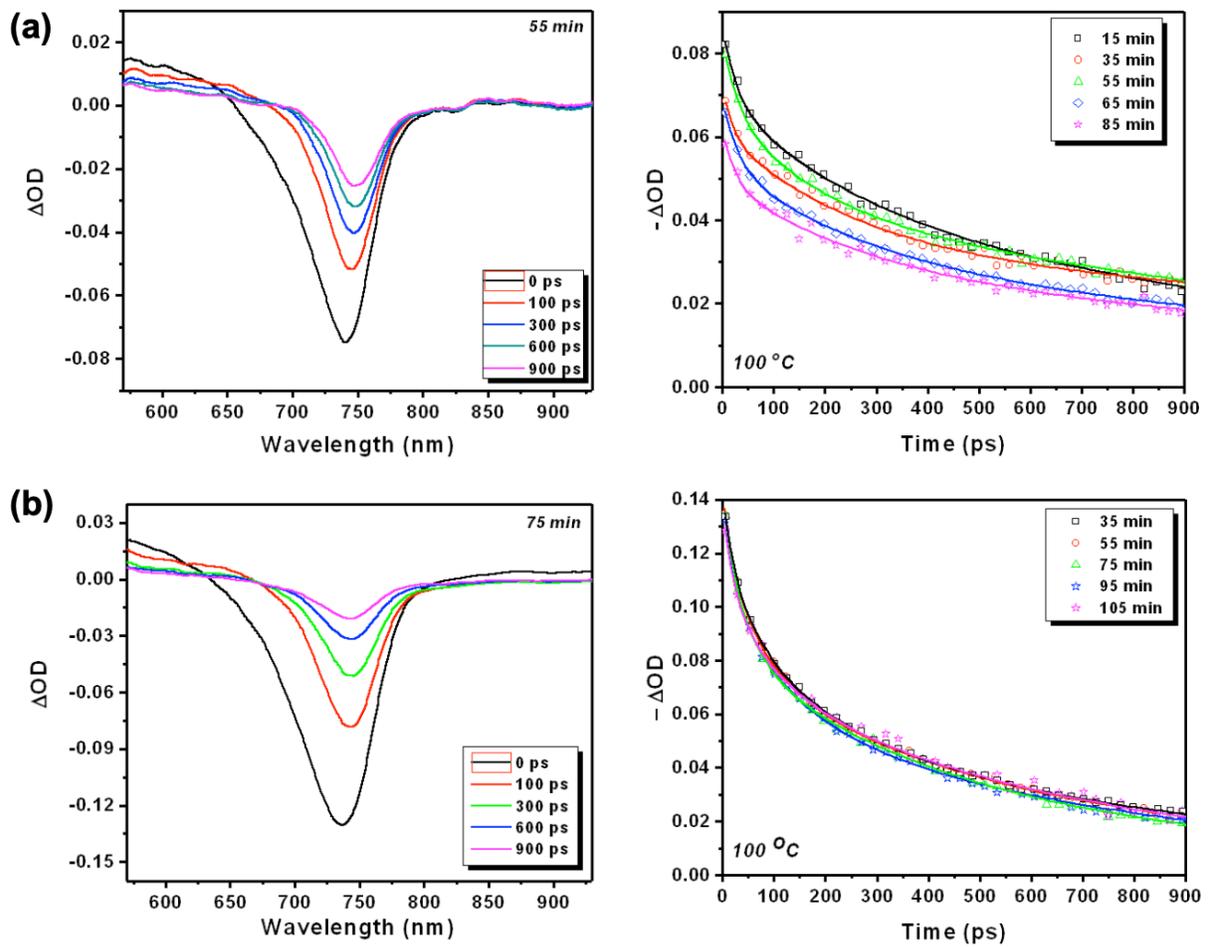

**Figure 4:** Typical TAS spectra of ΔOD as a function of wavelength and delay time of the PEDOT:PSS/CH$_3$NH$_3$PbI$_3$ (a) and PTAA/ CH$_3$NH$_3$PbI$_3$ (b) structures and corresponding ΔOD vs wavelength plots at various time delays following photoexcitation at 1026 nm with a pump fluence of 1.5 mJ cm$^{-2}$ at 100 ºC.

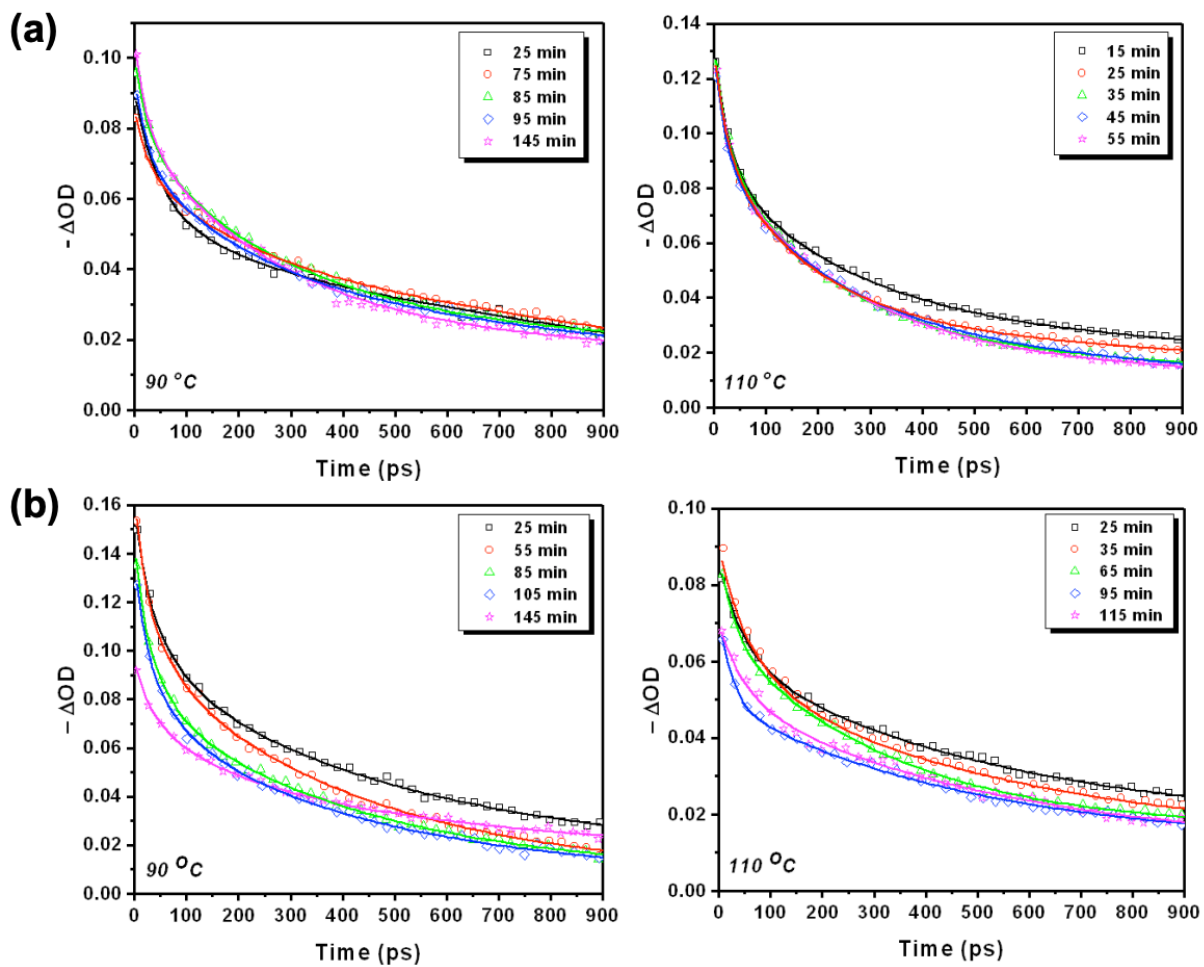

**Figure 5:** Transient band edge bleach kinetics (symbols) and the corresponding three-exponential fittings (lines) for the PEDOT:PSS/CH$_3$NH$_3$PbI$_3$ (a) and for PTAA/CH$_3$NH$_3$PbI$_3$ (b) architectures at 90 and 110 °C, photoexcited at 1026 nm with a pump fluence of 1.5 mJ cm$^{-2}$.

**Table 1:** HTL/CH$_3$NH$_3$PbI$_3$ Sample Configurations and Corresponding Synthesis Parameters.

| Samples | HTL polymers | Crystallization parameters | Annealing Temperature |
|---------|--------------|----------------------------|------------------------|
| 1 | PEDOT:PSS | Thermal Annealing | 100 °C |
| 2 | PEDOT:PSS | In-situ TAS Annealing | 90 °C |
| 3 | PEDOT:PSS | In-situ TAS Annealing | 100 °C |
| 4 | PEDOT:PSS | In-situ TAS Annealing | 110 oC |
| 5 | PTAA | Thermal Annealing | 100 °C |
| 6 | PTAA | In-situ TAS Annealing | 90 °C |
| 7 | PTAA | In-situ TAS Annealing | 100 °C |
| 8 | PTAA | In-situ TAS Annealing | 110 °C |

**Table 2:** Time Components for the PEDOT:PSS/CH$_3$NH$_3$PbI$_3$ Architectures Following Exponential Fitting ([a] Taken from (21)).

| Minutes | $\lambda_{max}$ (nm) | $\tau_1$ (ps) ± 2 ps | $\tau_2$ (ps) ± 8 ps | $\tau_3$ (ps) ± 0.5E7 ps |
|---|---|---|---|---|
| **[a]Thermal Annealing @ 100 °C** | | | | |
| 30 | 753 | 43 | 257 | 1.2E5 |
| **In-situ TAS Annealing @ 90 °C** | | | | |
| 25 | 738 | 45 | 185 | 2.7E7 |
| 75 | 737 | 25 | 210 | 1.7E7 |
| 85 | 738 | 22 | 105 | 3.2E7 |
| 95 | 739 | 25 | 235 | 8.6E6 |
| 145 | 737 | 25 | 276 | 2.1E7 |
| **In-situ TAS Annealing @ 100 °C** | | | | |
| 15 | 742 | 29 | 321 | 6.8E6 |
| 35 | 744 | 29 | 267 | 1.5E7 |
| 55 | 742 | 22 | 190 | 1.3E7 |
| 65 | 737 | 31 | 310 | 8.0E7 |
| 85 | 743 | 35 | 303 | 1.2E7 |
| **In-situ TAS Annealing @ 110 °C** | | | | |
| 15 | 731 | 29 | 257 | 1.5E7 |
| 25 | 731 | 22 | 210 | 8.7E6 |
| 35 | 731 | 24 | 225 | 1.9E7 |
| 45 | 731 | 24 | 230 | 1.7E7 |
| 55 | 731 | 24 | 257 | 1.4E7 |

**Table 3:** Time Components for the PTAA/CH$_3$NH$_3$PbI$_3$ Architectures Following Exponential Fitting ([a] Taken from (21)).

| Minutes | $\lambda_{max}$ (nm) | $\tau_1$ (ps) ± 2 ps | $\tau_2$ (ps) ± 8 ps | $\tau_3$ (ps) ± 13 ps |
|---|---|---|---|---|
| [a]**Thermal Annealing @ 100 °C** | | | | |
| 30 | 750 | 10 | 101 | 717 |
| **In-situ TAS Annealing @ 90 °C** | | | | |
| 25 | 741 | 21 | 69 | 506 |
| 55 | 737 | 18 | 53 | 418 |
| 85 | 732 | 17 | 48 | 418 |
| 105 | 731 | 21 | 74 | 444 |
| 145 | 739 | 29 | 110 | 540 |
| **In-situ TAS Annealing @ 100 °C** | | | | |
| 35 | 737 | 20 | 85 | 508 |
| 55 | 736 | 22 | 83 | 574 |
| 75 | 734 | 21 | 60 | 504 |
| 95 | 734 | 21 | 124 | 833 |
| 105 | 740 | 22 | 102 | 660 |
| **In-situ TAS Annealing @ 110 °C** | | | | |
| 25 | 736 | 50 | 522 | 522 |
| 35 | 737 | 19 | 63 | 429 |
| 65 | 736 | 30 | 338 | 338 |
| 95 | 739 | 26 | 503 | 503 |
| 115 | 739 | 51 | 510 | 510 |

**Table 4:** Recombination Rate Constants for the PEDOT:PSS/$CH_3NH_3PbI_3$ Architectures Following Polynomial Fitting ([a] Taken from (21)).

| Minutes | $\lambda_{max}$ (nm) | $k_3$ ($cm^6s^{-1}$) ± 0.4 | $k_2$ ($cm^3s^{-1}$) ± 0.2 | $k_1$ ($\mu s^{-1}$) ± 0.1 |
|---|---|---|---|---|
| [a]**Thermal Annealing @ 100 °C** | | | | |
| 30 | 753 | 7.3 x $10^{-14}$ | 3.6 x $10^{-11}$ | 4.4 x $10^{-6}$ |
| **In-situ TAS Annealing @ 90 °C** | | | | |
| 85 | 738 | 7.4 x $10^{-16}$ | 2.4 x $10^{-12}$ | 1.2 x $10^{-7}$ |
| **In-situ TAS Annealing @ 100 °C** | | | | |
| 55 | 742 | 4.5 x $10^{-16}$ | 3.3 x $10^{-12}$ | 8.7 x $10^{-7}$ |
| **In-situ TAS Annealing @ 110 °C** | | | | |
| 25 | 731 | 1.2 x $10^{-15}$ | 3.6 x $10^{-12}$ | 2.2 x $10^{-7}$ |

**Table 5:** Recombination Rate Constants for the PTAA/$CH_3NH_3PbI_3$ Architectures Following Polynomial Fitting ([a] Taken from (21)).

| Minutes | $\lambda_{max}$ (nm) | $k_3$ ($cm^6s^{-1}$) ± 0.4 | $k_2$ ($cm^3s^{-1}$) ± 0.2 | $k_1$ ($\mu s^{-1}$) ± 0.1 |
|---|---|---|---|---|
| [a]**Thermal Annealing @ 100 °C** | | | | |
| 30 | 750 | 3.8 x $10^{-14}$ | 1.7 x $10^{-11}$ | 3.1 x $10^{-6}$ |
| **In-situ TAS Annealing @ 90 °C** | | | | |
| 85 | 732 | 1.8 x $10^{-15}$ | 2.2 x $10^{-12}$ | 1.8 x $10^{-6}$ |
| **In-situ TAS Annealing @ 100 °C** | | | | |
| 75 | 734 | 1.7 x $10^{-15}$ | 4.8 x $10^{-12}$ | 1.7 x $10^{-6}$ |
| **In-situ TAS Annealing @ 110 °C** | | | | |
| 35 | 737 | 3.4 x $10^{-15}$ | 4.1 x $10^{-12}$ | 3.4 x $10^{-6}$ |

**TOC Graphic:**

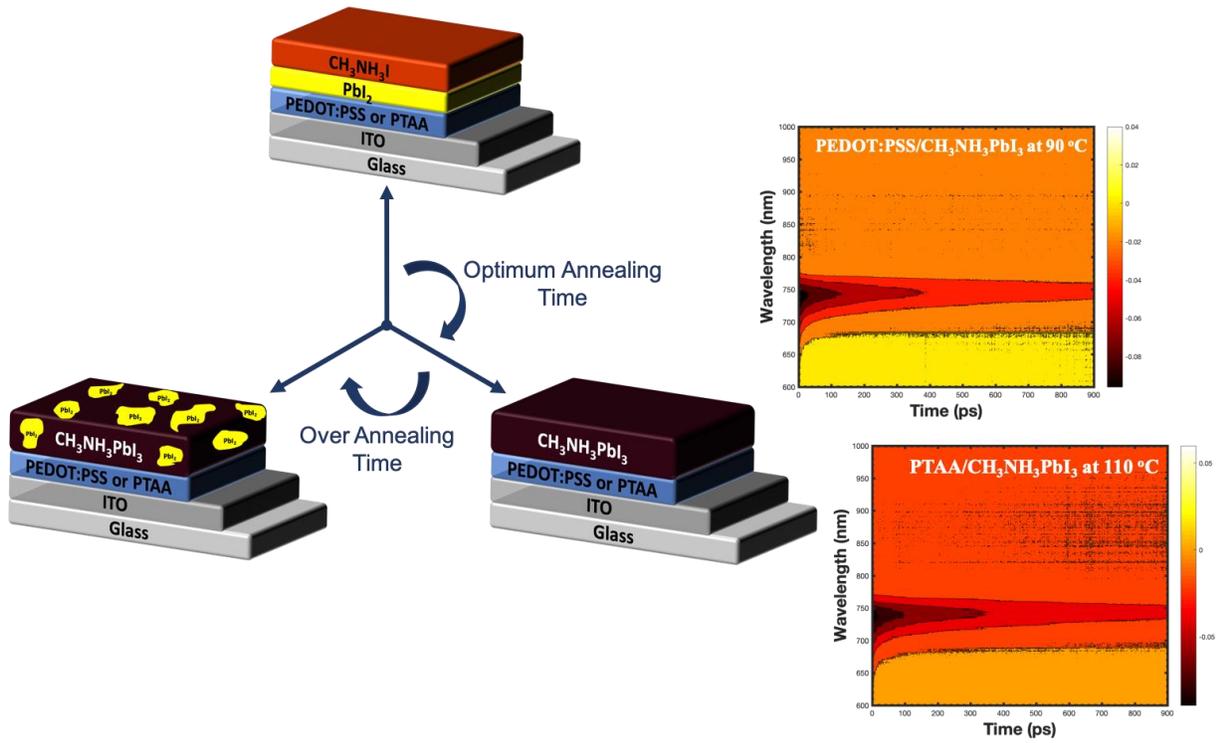